\pdfoutput=1
\documentclass{aa}

\usepackage{graphicx}
\usepackage{txfonts}
\usepackage{booktabs}
\let\linenumbers\nolinenumbers

\newcommand{\svt}{SVT}

\begin{document}

\title{A transformer analysis lowers the energy threshold and improves the
       sensitivity of a compact imaging atmospheric Cherenkov telescope}
\titlerunning{A transformer analysis of a compact IACT}
\authorrunning{E. Jobst et al.}

\author{E. Jobst\inst{1,2} \and D. Paneque\inst{1}\corrauth{dpaneque@mpp.mpg.de}
        \and L. Heinrich\inst{2}}
\institute{Max-Planck-Institut f\"ur Physik (Werner-Heisenberg-Institut),
           Boltzmannstr.\ 8, 85748 Garching, Germany
      \and Technical University of Munich, School of Natural Sciences,
           85748 Garching, Germany}

\date{Received ---; accepted ---}

\abstract
{A companion study simulated an idealised stand-alone imaging atmospheric Cherenkov
 telescope (IACT) with a 5\,m diameter mirror. A video vision transformer that processes the raw camera
 waveforms lowered its energy threshold by a factor of three, compared with the
 standard analysis based on image parameters.}
{We use realistic photodetectors, compute the differential flux sensitivity, and
 study how hardware parameters affect the transformer analysis.}
{With the simulation tools of the Cherenkov Telescope Array Observatory, we
 compare three photodetector efficiencies (100\,\%, 50\,\% and that of the LST-1
 photomultipliers) with the transformer and with a standard analysis optimised for each.
 We then vary the dish diameter, field of view, pulse width and sampling rate.}
{With the LST-1 photomultipliers, the energy threshold is 0.18\,TeV with the
 transformer and 0.73\,TeV with the standard analysis. The standard analysis
 needs a photodetector efficiency of 100\,\% to match that threshold.
 The main reason is that the transformer works with events of about 35 photoelectrons
 in the camera, whereas the standard analysis needs about 200 before image cleaning. A 4\,m dish with the transformer is more
 sensitive at all energies than a 5\,m dish with the standard analysis. Dishes of
 4, 5 and 6\,m diameter give thresholds of 0.38, 0.18 and 0.13\,TeV. A
 $4^{\circ}$ field-of-view camera is 1.2--1.8 times less sensitive than a $6^{\circ}$ camera.
 Sampling the pixel signals at 2\,GS/s instead of 5\,GS/s raises the threshold to 0.26\,TeV, whereas
 the single-photoelectron pulse width (1--3\,ns) matters little.}
{The analysis matters as much as the mirror area and the photodetector efficiency.
 With the transformer analysis, the idealised 5\,m diameter telescope comes within a
 factor of about four of the sensitivity of LST-1 and MAGIC above 0.25\,TeV, although
 its mirror area is 21 times smaller than that of the 23\,m LST-1. This performance
 should improve substantially in the coming years.}


\maketitle

\section{Introduction}
\label{sec:intro}

For the last forty years, the reconstruction of events recorded by imaging atmospheric
Cherenkov telescopes (IACTs) has been mostly based on the parameterisation of the
shower image introduced by \citet{hillas1985}. Some IACTs take a snapshot of the
Cherenkov flash, whereas many others record a short movie of it, with a waveform in
every pixel. Each waveform is converted into one charge and one arrival time. The pixels
that probably contain only night-sky light are then removed, and the image is described
by a few image parameters. The event is reconstructed from these parameters, nowadays
often with machine-learning methods such as random forests. Each of the 3600
pixels of the camera considered here records 128 samples, so an event contains about
$4.6\times10^{5}$ values. These values are reduced to about ten parameters before
any reconstruction begins \citep{jobst2026}. This
approach is robust, well understood and used by most current instruments. We call
it the standard analysis. Its performance improves when more photons are detected
from each shower, with larger mirrors or better photodetectors. It also improves with
stereoscopic observations, in which several telescopes observe the same shower.

Other methods exist. The CAT telescope compared each image with a semi-analytical model
of the shower \citep{lebohec1998}, a method that H.E.S.S. developed further
\citep{denaurois2009}. Other analyses fit templates of simulated images to the data
\citep{parsons2014}, or apply deep learning to the images
\citep{shilon2019,miener2024}. Most deep-learning studies use convolutional neural
networks (CNNs). They replace one or more stages of the standard analysis, not the
whole chain, and they work on calibrated images, usually after image cleaning
\citep{jobst2026}. With such a network, \citet{jacquemont2021} reconstructed the real
data of LST-1, the first Large-Sized Telescope of the Cherenkov Telescope Array
Observatory (CTAO), better than the standard analysis. Networks trained on simulations
have also been adapted to the real data of LST-1 \citep{dellaiera2026}.
\citet{miener2025} applied a CNN to the simulated waveforms of LST-1, but still after
the standard image cleaning. These networks are designed for existing telescopes,
which were themselves designed for the standard analysis. Their hexagonal pixels, for
example, do not fit the square grid that networks developed for images expect.
Transformers have been applied to simulations of a future array of water Cherenkov
detectors for gamma rays, and they outperformed the established reconstruction
methods \citep{pirke2026}. To our knowledge, \citet[][hereafter Paper~1]{jobst2026}
were the first to apply a video vision transformer to IACT data, and the first to
analyse the raw waveforms, with neither calibration nor image cleaning.

Paper~1 showed that the low-energy limit of a
compact telescope depends at least as much on the analysis as on the mirror area.
Paper~1 simulated the Small Virtual Telescope (\svt), a deliberately idealised stand-alone 5\,m diameter telescope, which observes in monoscopic
(mono) mode. It compared two analyses of the same simulated events: a video vision
transformer \citep{vaswani2017,dosovitskiy2021,arnab2021}, which processes the raw waveform movie, and the standard analysis, re-optimised for the same telescope. We call the first
one the transformer analysis. Compared with the standard analysis, it lowered the
energy threshold by a factor of three,
reconstructed the direction of the gamma rays down to energies four times lower,
and raised the effective area
at 0.2\,TeV by a factor of 3.3. Paper~1 did not report the flux sensitivity of the
telescope. This paper reports it, with realistic photodetectors, and studies how the
performance of the transformer analysis depends on a few hardware parameters of the
telescope.

We take two steps. First (Sect.~\ref{sec:qe}), we replace the photodetectors of
Paper~1, which have a quantum efficiency (QE) of 100\,\%, with realistic ones. We follow the performance
of both analyses as the number of detected photons decreases, compute the differential
flux sensitivity, and compare it with the sensitivities of LST-1 and MAGIC.
Second (Sect.~\ref{sec:scan}), we study how the performance of the transformer
depends on the hardware. We vary the dish diameter and the field of view of the camera one at a time, and the
width of the photodetector pulse and the sampling rate of the waveforms together. The scientific motivation for such a telescope and
for an array of them is discussed in a companion paper \citep[][hereafter
Paper~3]{paper3}.

\section{Simulations and analyses}
\label{sec:sims}

\subsection{Simulation chain}
\label{sec:mc}

Air showers are simulated with CORSIKA~7.7550 \citep{heck1998}, and the telescope
response with \texttt{sim\_telarray} version 2025.51.0 \citep{bernlohr2008}. Both
are configured with the CTAO \texttt{simtools} package \citep{maier2026simtools}. We
therefore use the same tools as CTAO. Gamma rays are simulated between
10\,GeV and 10\,TeV, and protons between 30\,GeV and 30\,TeV. The two energy
ranges are shifted because, in this energy range, a proton produces an image of
similar brightness to
that of a gamma ray of about a third of its energy. For simplicity, the background
consists of protons only, the dominant component of the cosmic rays. Helium and
heavier nuclei, and cosmic-ray electrons, are neglected (Sect.~\ref{sec:realism}). The telescope is placed at the
CTAO-North site on La Palma, at 2200\,m, and points at a zenith angle of
$10^{\circ}$. The training samples of gamma rays and protons are diffuse: they arrive within
$4.25^{\circ}$ of the pointing direction, that is, in a cone of $8.5^{\circ}$ in
diameter. This cone extends more than $1^{\circ}$ beyond the sides of the largest
camera studied. For the test samples, the gamma rays come from a point source at the centre of
the camera, and the protons are diffuse in the same cone. The protons are
weighted to follow the spectrum measured by BESS (index $-2.7$; \citealt{sanuki2000}).
The gamma rays are weighted to follow the spectrum of the Crab Nebula measured by
MAGIC \citep[Eq.~1]{aleksic2016}. As in Paper~1, the background in the region around the source, the ON region, is
estimated from source-free OFF regions within
$0.5^{\circ}$ of the centre of the camera (Sect.~\ref{sec:fom}).

\paragraph{Differences from Paper~1.} Four changes affect the comparison with the
results of Paper~1. First, Paper~1 simulated gamma rays and protons within
$2.5^{\circ}$ of the pointing direction. The wider cone used here is more realistic.
Protons that arrive up to $4.25^{\circ}$ from the centre of the camera also produce
images in it, and some of them resemble gamma rays. A smaller fraction of the simulated protons then triggers the telescope, so more
protons must be simulated. The network must therefore also learn to reject a more varied
background, which is a more difficult task. Second, Paper~1 weighted the gamma rays to the
spectrum measured by HEGRA, a power law of photon index 2.6
\citep{aharonian2004crab}. The curved spectrum of \citet{aleksic2016} is harder below
1\,TeV, which moves the energy threshold to higher energies (Sect.~\ref{sec:fom}).
These two changes make the results somewhat worse than those that the configuration of
Paper~1 would give, in particular at the lowest energies. The effect is largest when
fewer Cherenkov photons are detected, as with the photomultipliers of LST-1
(Sect.~\ref{sec:qe}). With the simulation settings of Paper~1, which we used for our
preliminary results of 2026, the sensitivity of the 5\,m telescope with these
photomultipliers was 1.8 times better at 0.3\,TeV and 1.2 times better at 1.3\,TeV.
Third, Paper~1 estimated the background from three OFF regions,
whereas all the analyses of this paper use five. With more OFF regions, the
background is estimated more precisely, which improves the sensitivity by up to
5\,\% where the background dominates. Fourth, the transformer gives each event a gamma/hadron
score, which is high for events that look like gamma rays (Sect.~\ref{sec:vivit}). To
keep more gamma rays at the lowest energies, the cut on this score is lowered from
0.9 to 0.8 below 0.16\,TeV (Sect.~\ref{sec:fom}). This change affects only the bins below 0.16\,TeV.
Section~\ref{sec:conclusions} discusses other changes of the analysis, which we expect
to improve these results.

A single set of CORSIKA showers is used for every configuration of
Sects.~\ref{sec:qe} and \ref{sec:scan}. Only the \texttt{sim\_telarray} stage, which
is fast, is repeated. The shower-to-shower fluctuations are therefore the same for
all configurations, so that differences between them are due only to the hardware
and to the training. This also saves computing time. The core-scatter radius, the largest simulated distance
between the telescope and the shower axis, is 250\,m for both gamma rays and
protons, as in Paper~1.

\subsection{Baseline telescope}
\label{sec:baseline}

The baseline is the \svt\ of Paper~1 (Table~\ref{tab:svt}): a 5\,m parabolic dish
with a focal ratio of 1.5, a square camera of $60\times60$ pixels of
$0.1^{\circ}$ covering $6.1^{\circ}$ on a side, a single-photoelectron pulse of 1\,ns full
width at half maximum (FWHM), and waveform sampling at 5\,GS/s over 128 samples.
In Sect.~\ref{sec:qe}, we analyse it with three photodetector efficiencies: 100\,\%,
50\,\% and that of the photomultipliers of LST-1, which the rest of the paper
uses. In Sect.~\ref{sec:scan}, we vary the dish diameter, the field of view, and
the pulse width and sampling rate. The other idealisations of Paper~1 are kept
(Sect.~\ref{sec:realism}).

\begin{table}[t]
\caption{Baseline telescope, the Small Virtual Telescope (\svt) of Paper~1, and the
changes made in this paper.}
\label{tab:svt}
\centering
\small\setlength{\tabcolsep}{3pt}
\begin{tabular}{l l l}
\toprule
Parameter & Paper~1 & This paper \\
\midrule
Dish diameter         & 5\,m                        & 4, 5, 6\,m (Sect.~\ref{sec:scan_dish}) \\
Focal ratio $f/D$     & 1.5                         & 1.5 \\
Mirror reflectivity   & 100\,\%                     & 100\,\%\tablefootmark{a} \\
Mirror alignment      & perfect                     & perfect\tablefootmark{a} \\
Pixel size            & $0.1^{\circ}$               & $0.1^{\circ}$ \\
Field of view         & $6^{\circ}$                 & $4^{\circ}$, $6^{\circ}$ (Sect.~\ref{sec:scan_fov}) \\
\multicolumn{3}{l}{Photodetector} \\
\quad Quantum efficiency & 100\,\%                  & 100, 50\,\%, LST-1 \\
\quad Single-phe pulse (FWHM) & 1\,ns               & 1, 2, 3\,ns \\
\quad Transit-time spread & 0                       & 0\tablefootmark{a} \\
Sampling              & 5\,GS/s                     & 5, 2\,GS/s \\
Electronic noise      & none                        & none\tablefootmark{a} \\
Night-sky background  & 0.15\,GHz/pixel             & dark sky, $\propto D^{2}\theta_{\rm pix}^{2}$\tablefootmark{a} \\
Background particles  & protons                     & protons\tablefootmark{a} \\
Trigger               & analogue sum\tablefootmark{b} & as in Paper~1 \\
\bottomrule
\end{tabular}
\tablefoot{phe: photoelectrons. \tablefoottext{a}{Idealisation of Paper~1, kept in
this paper (Sect.~\ref{sec:realism}).} \tablefoottext{b}{The signals are summed over regions of
$5\times5$ pixels ($0.5^{\circ}$), which overlap by half a region. The telescope
triggers when a sum exceeds 10\,phe.}}
\end{table}

\subsection{Remaining idealisations}
\label{sec:realism}

Paper~1 listed the idealisations that a realistic study must address: those of the
night-sky background, the optics, the quantum efficiency, the transit-time spread,
the electronic noise and the pulse shape. This paper removes one of them, the QE of
100\,\% (Sect.~\ref{sec:qe}), and varies the pulse width
(Sect.~\ref{sec:scan_pulse}). The others are kept (Table~\ref{tab:svt}). The night-sky background is that of a dark sky. The mirror
reflects all the light, and it has neither facet errors nor misalignment. The dish is
parabolic, so that the photons from a source at the centre of the field of view reach
the camera at the same time. Other designs, such as Davies--Cotton dishes, spread these arrival
times. A parabolic dish also spreads them, but only for light from far off the
centre. The photodetector pulse has no transit-time spread, that is, no random variation of
its delay, and the electronics add no noise.
The background consists of protons only (Sect.~\ref{sec:mc}). These protons
arrive only within $4.25^{\circ}$ of the pointing direction, and their shower axes pass
only up to 250\,m from the telescope. Protons from larger angles or distances can
also produce light in the camera. In every bin of estimated energy, the protons that
survive the cuts come from all distances up to 250\,m, and their number per unit
distance decreases only slowly towards this limit. The simulated background is
therefore probably a few tens of per cent too low. The sensitivity is affected less.
Where the background dominates, the weakest detectable flux grows only as the square
root of the background. A background that is 20\,\% too low, for example, makes the
weakest detectable flux only about 10\,\% too low.
The low background affects both analyses of the \svt, not only the transformer. Both
analyses use the same simulated events, so the comparison between them is fair. The
low background matters mainly for the comparison with MAGIC and LST-1, whose
sensitivities come from real data.
Each of these idealisations favours the simulated telescope, so the absolute sensitivities of this
paper are optimistic. Their effect will be studied in a future paper
(Sect.~\ref{sec:conclusions}).

\subsection{The transformer analysis}
\label{sec:vivit}

The network is the factorised-encoder video vision transformer (ViViT-FE) of
Paper~1, with about $7\times10^{6}$ parameters. The waveform movie is cut into small blocks, called tubelets, of $10\times10$ pixels
($1^{\circ}\times1^{\circ}$) and two time samples. Each tubelet is converted into a vector,
called a token. A spatial encoder combines the tokens of each pair of time samples
into one vector, and a temporal encoder then processes the sequence of these
vectors. A single output layer returns the energy, the
two coordinates of the arrival direction in the camera plane, and a gamma/hadron
score, which is high for events that look like gamma rays. The network learns all
these quantities at once, with a loss function that balances them through gradient
normalisation.

The same network is used for every configuration of Sects.~\ref{sec:qe} and
\ref{sec:scan}. It has the same architecture and the same hyperparameters, that is,
the same settings of the network and of its training, such as the learning rate. It
is trained separately for each configuration. This keeps the study simple, and ensures that differences between configurations are
not caused by changes in the analysis. It also limits the configurations that can be
studied. The configurations of this paper either reduce the information recorded by
the camera (a smaller field of view, a wider pulse or a lower sampling rate), or
change the number of detected photons (another QE or dish diameter). They do not change
the structure of the recorded images, so that the network designed for the baseline
can process them. Other pixel sizes, or larger fields of view, would require a
network adapted to them, with more parameters (Sect.~\ref{sec:conclusions}).
The images of the $4^{\circ}$ camera and the 52 samples recorded at 2\,GS/s
(Sect.~\ref{sec:scan_pulse}) enter the network without any change of its structure.
Both are simpler than the baseline, so they add no complexity. The number of
simulated events is also large enough for the training to converge with this number
of parameters. The hyperparameters are those of Paper~1,
where they were not optimised for any telescope, the baseline included. Their
optimisation for each configuration is left for future work
(Sect.~\ref{sec:conclusions}).

To test the stability of the training, the baseline with photodetectors of 100\,\%
QE was trained four times, with different random seeds. This test was made with an earlier
version of the simulations, and we expect the same result with the final one. Above
0.1\,TeV, the four sensitivities differ by at most 12\,\%, and their scatter (rms) is
1.5--5\,\%, smaller than their statistical uncertainties. This scatter comes from
the training of the network, which was repeated on the same simulated events. We
expect it to be similar for every configuration, and we therefore do not interpret
differences of this size between configurations.

\subsection{The reference analysis}
\label{sec:std}

The reference is the LST-1 analysis chain \texttt{lstchain} v0.10.5
\citep{lopezcoto2022}. It performs calibration by peak search and charge
integration, image cleaning, Hillas parameterisation, and random forests for
energy, direction and gamma/hadron separation. As in Paper~1, every stage is
re-optimised for the telescope to which it is applied, with \texttt{optuna}
\citep{akiba2019}, using a tree-structured Parzen estimator and 150 trials per stage.
A standard analysis tuned for a different telescope would not be a fair reference.
Re-optimising the standard analysis for a new telescope takes far longer than
training the transformer for it, and Sect.~\ref{sec:scan} concerns the transformer.
The standard analysis is therefore run, fully re-optimised, only on the
photodetector configurations of Sect.~\ref{sec:qe}, where it provides the reference.

\subsection{Figures of merit}
\label{sec:fom}

\begin{figure*}[t]
\centering
\includegraphics[width=\textwidth]{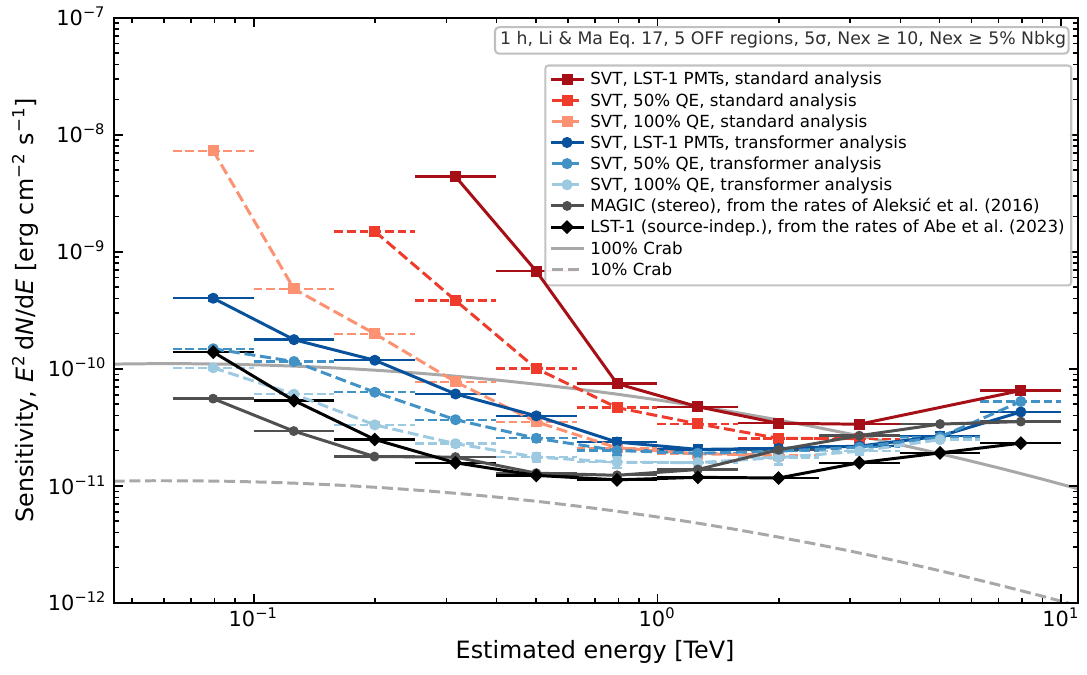}
\caption{Differential point-source sensitivity in one hour of the 5\,m baseline
telescope for three photodetector responses (100\,\% and 50\,\% QE, and the PMTs of
LST-1), analysed with the standard analysis (red) and with the transformer (blue),
compared with LST-1 and MAGIC. Lower curves mean better sensitivity: each point gives
the weakest flux of a source that can be detected in one hour. Solid lines: PMTs of
LST-1; dashed lines: 50\,\% and 100\,\% QE. The curves of LST-1 and MAGIC are computed from their rates of gamma rays and background, for 1\,h
(Sect.~\ref{sec:fom} and Appendix~\ref{app:ref}). The grey lines show
100\,\% and 10\,\% of the flux of the Crab Nebula. The error bars, which are small
and mostly hidden by the markers, show the statistical uncertainties due to the
number of simulated events. The simulated
telescope is still idealised (Sect.~\ref{sec:realism}).}
\label{fig:qe}
\end{figure*}

\paragraph{Differential sensitivity.} The main figure of merit is the
differential point-source flux sensitivity. It is the minimum flux, in each of five
bins per decade of estimated energy, that gives a $5\sigma$ excess according to
Eq.~(17) of \citet{lima1983}, with the background estimated from five OFF regions
(Sect.~\ref{sec:mc}). We also require at least ten excess events, and an excess of
at least 5\,\% of the residual background. All the sensitivities of this paper are
for an exposure of 1\,h. Two cuts select the events in each energy bin: a cut on the
gamma/hadron score and a cut on the angular distance to the source. For the standard
analysis, the cut on the score keeps 80\,\% of the gamma rays in each bin of
estimated energy. For the transformer analysis, the cut on the score is fixed, not
varied in each energy bin: it requires a score above 0.9 at energies above
0.16\,TeV, and above 0.8 at lower energies, where this looser cut keeps more gamma
rays (Sect.~\ref{sec:mc}). For both analyses, the cut on the
angular distance then keeps 68\,\% of the gamma rays that pass the cut on the score.
The cuts could be optimised in each energy bin. We use instead simple cuts because
our aim is a generic comparison of the analysis results for different telescope
configurations. A dedicated optimisation would improve the sensitivities by up to a few
tens of per cent, but the result would depend on how the cuts are optimised in each
bin and for each configuration. An energy bin is used only if at least ten
simulated gamma rays and four simulated protons survive the cuts. At multi-TeV
energies, the separation of gamma rays from hadrons is very good. Most simulated
protons are removed, the signal is almost free of background, and the sensitivity
is limited by the number of gamma rays. The minimum number of simulated protons is
therefore only four, so that the sensitivity can be given at these energies. The
statistical uncertainties of the sensitivity come from the number of simulated
events. Configurations are compared through their sensitivity in each energy bin
and through their energy threshold.

\paragraph{Energy threshold.} We define the energy threshold as the peak, in true
energy, of the distribution of the simulated gamma rays that survive all cuts,
weighted to the spectrum of the Crab Nebula of \citet[Eq.~1]{aleksic2016}. This
spectrum is curved: its local photon index is about 2.1 at 0.1\,TeV and 2.3 at
0.3\,TeV. The threshold is not the lowest energy at which a sensitivity can be
given. Below the peak, the rate of gamma rays falls but does not vanish, so the
sensitivity can be computed in lower bins as long as enough simulated events survive
the cuts (see above). The sensitivity curve therefore starts below the threshold
(Sect.~\ref{sec:qe}).

\paragraph{Reference instruments.} We compare the \svt\ with LST-1, the first
Large-Sized Telescope of CTAO, a 23\,m telescope observing in mono
\citep{abe2023}, and with the two 17\,m MAGIC telescopes observing in stereo
\citep{aleksic2016}. The two collaborations report the sensitivity for 50\,h of
observation. This time is not relevant for transient sources, such as flares of
active galactic nuclei, novae and gamma-ray bursts, which are the main scientific
targets of the \svt\ (Paper~3). We therefore use the sensitivities of MAGIC and LST-1
in 1\,h, as for the \svt. We compute them from the rates of gamma rays and background
after cuts, with the same criteria as for the \svt. For MAGIC, the rates are those
published in the appendix of \citet{aleksic2016}, measured at zenith angles below
$30^{\circ}$. For LST-1, they are the rates of the source-independent analysis of
\citet{abe2023} with cuts optimised for 1\,h, provided by the LST collaboration.
Appendix~\ref{app:ref} explains how these sensitivities are derived, and compares
them with the published curves for 50\,h. The sensitivities in Crab units are
converted to erg\,cm$^{-2}$\,s$^{-1}$ with the spectrum of the Crab Nebula of
\citet[Eq.~1]{aleksic2016}. The sensitivities of the \svt\ come from simulations,
which are still simple and idealised (Sect.~\ref{sec:realism}), whereas those of MAGIC
and LST-1 come from real observations of the Crab Nebula. The curves are therefore
not fully comparable. The comparison serves to put the sensitivity of the \svt\ with
the transformer analysis into context. We compare the \svt\ with MAGIC and LST-1
only up to 6.3\,TeV. The simulated gamma rays end at 10\,TeV (Sect.~\ref{sec:mc}).
The last bin, 6.3--10\,TeV, therefore misses the gamma rays of higher energy that
would be reconstructed in it, and the sensitivity of the \svt\ in this bin is
somewhat pessimistic.

\section{Photodetector efficiency: from an ideal QE of 100\,\% to the PMTs used in LST-1}
\label{sec:qe}

Paper~1 used a QE of 100\,\% between 260 and 650\,nm. To see how each analysis depends on the number of detected photons,
we analyse the baseline telescope with three photodetector responses. The first
is the QE of 100\,\% of Paper~1, and the second a uniform QE of 50\,\%. The third is the
measured, wavelength-dependent QE of the photomultipliers (PMTs) installed in LST-1
\citep{abe2023}, which were produced in 2015 and have been in operation since
2018. Only this last response is realistic. It gives the 5\,m telescope the QE of
LST-1. Each
configuration is analysed with the transformer and with the standard analysis,
re-optimised for each. Figure~\ref{fig:qe} shows the differential sensitivities, and
Table~\ref{tab:qe} gives selected values. The
transformer curve with the LST-1 PMTs is the baseline curve of Sect.~\ref{sec:scan}.

\subsection{Sensitivity}
\label{sec:qe_sens}

\paragraph{The transformer is less affected by a loss of photons.} Going from
photodetectors of 100\,\% QE to the LST-1 PMTs, the energy threshold rises from 0.074
to 0.18\,TeV with the transformer, and from 0.19 to 0.73\,TeV with the standard
analysis. The threshold thus rises by a factor of 2.4 with the transformer and of 3.8
with the standard analysis. The difference is much larger for the sensitivity at a
given energy. From 100\,\% QE to the LST-1 PMTs, the sensitivity of the transformer
analysis degrades by a factor of 2.7
at 0.32\,TeV, 2.2 at 0.5\,TeV and 1.5 at 0.8\,TeV. That of the standard analysis
degrades by factors of 19 at 0.5\,TeV and 3.5 at 0.8\,TeV. With
the LST-1 PMTs, the standard analysis reaches no bin below 0.25\,TeV,
whereas the transformer analysis reaches the bin of 0.063--0.1\,TeV. Paper~1
proposed the following explanation. The image cleaning of the standard analysis keeps only the pixels whose charge
exceeds a fixed threshold. Each lost photon can push a pixel below this threshold, and so
shrinks the image on which the Hillas parameters are computed. The transformer, instead, uses the whole space-time pattern of the
signal, including the arrival times of photons that no cleaning would keep.
Section~\ref{sec:qe_pe} quantifies how much of the signal the standard analysis
discards.

\paragraph{The transformer is worth more than a 100\,\% QE photodetector.} With the
LST-1 PMTs, the transformer analysis is more sensitive than the standard analysis
with photodetectors of 100\,\% QE by a factor of 2.7 at 0.13\,TeV, 1.7 at 0.2\,TeV
and 1.3 at 0.32\,TeV. In the lowest bin, 0.063--0.1\,TeV, far below the threshold of
the standard analysis, the factor is $18\pm1$. Between
0.4 and 2.5\,TeV, the standard analysis with 100\,\% QE is about 10\,\% more
sensitive, a difference comparable with the statistical uncertainties. Above
2.5\,TeV, fewer than four simulated protons survive its cuts, so that it gives no
sensitivity there (Sect.~\ref{sec:fom}). The two have about the same energy
threshold, 0.18 and 0.19\,TeV. At the lowest energies, the transformer analysis with
the LST-1 PMTs thus gives an improvement that no increase of the QE could give, not
even to the theoretical maximum of 100\,\%.

\begin{table}[t]
\caption{Selected characteristics of the curves of Fig.~\ref{fig:qe}: energy
threshold, and differential sensitivity in one hour in two bins of estimated energy,
in Crab units (C.U.).}
\label{tab:qe}
\centering
\small\setlength{\tabcolsep}{4pt}
\begin{tabular}{l c c c}
\toprule
 & $E_{\rm th}$ & \multicolumn{2}{c}{Sensitivity [C.U.]} \\
 & [TeV] & 0.25--0.4\,TeV & 1.0--1.6\,TeV \\
\midrule
\multicolumn{4}{l}{\svt, standard analysis} \\
\quad LST-1 PMTs & 0.73 & $50.7\pm1.9$ & $0.99\pm0.05$ \\
\quad 50\,\% QE & 0.45 & $4.43\pm0.05$ & $0.71\pm0.04$ \\
\quad 100\,\% QE & 0.19 & $0.89\pm0.02$ & $0.39\pm0.04$ \\
\multicolumn{4}{l}{\svt, transformer analysis} \\
\quad LST-1 PMTs & 0.18 & $0.70\pm0.03$ & $0.43\pm0.03$ \\
\quad 50\,\% QE & 0.11 & $0.42\pm0.02$ & $0.40\pm0.03$ \\
\quad 100\,\% QE & 0.074 & $0.27\pm0.02$ & $0.33\pm0.03$ \\
\midrule
MAGIC & & 0.20 & 0.29 \\
LST-1 & & 0.18 & 0.25 \\
\bottomrule
\end{tabular}
\tablefoot{The uncertainties come from the number of simulated events. MAGIC and
LST-1 have no uncertainties (Appendix~\ref{app:ref}).}
\end{table}

\paragraph{The gain grows when fewer photons are detected.} Compared with the
standard analysis on the same events, the transformer improves the sensitivity, with
photodetectors of 100\,\% QE, by a factor of 6.0 at 0.2\,TeV, 3.4 at 0.32\,TeV, 2.0 at
0.5\,TeV and 1.3 at 0.8\,TeV, and it lowers the energy threshold by a factor of 2.6.
With 50\,\% QE, the factors are 10 at 0.32\,TeV, 3.9 at 0.5\,TeV and 2.3 at
0.8\,TeV, and the energy threshold is four times lower, 0.11 instead of 0.45\,TeV. With
the LST-1 PMTs, the factors are 17 at 0.5\,TeV and 3.2 at 0.8\,TeV, and
the energy threshold is again four times lower, 0.18 instead of 0.73\,TeV. For both
analyses, the curves with 50\,\% QE lie between the two other cases below
6.3\,TeV (Fig.~\ref{fig:qe}). This agrees with the cross-check of Paper~1 at half
efficiency, in which the transformer was also less affected than the standard
analysis.

\paragraph{Comparison with MAGIC and LST-1.} With the LST-1 PMTs, the \svt\
analysed with the transformer is 6.0--6.7 times less sensitive than MAGIC in stereo
between 0.1 and 0.25\,TeV, 3.1--3.5 times between 0.25 and 0.63\,TeV, and 1.5--1.9
times between 0.63 and 1.6\,TeV. It is as sensitive as MAGIC between 1.6 and
2.5\,TeV, and 1.2--1.3 times more sensitive between 2.5 and 6.3\,TeV, partly thanks to the larger field of view of its camera. Compared with LST-1, it is 3.2--4.8 times
less sensitive between 0.1 and 0.63\,TeV, and 1.4--2.1 times between 0.63 and
6.3\,TeV. This
compares a 5\,m telescope with a 23\,m telescope, both observing in mono, with the
same photodetector QE. The effective mirror area of the \svt,
17.65\,m$^{2}$ after camera shadowing, is 21 times smaller than that of LST-1, about
370\,m$^{2}$ \citep{hayashida2015}. The two are not yet compared on equal terms. The LST-1 curve comes from real data. The simulated \svt, instead, still has a perfectly
reflecting and aligned mirror, a 1\,ns photodetector pulse without transit-time
spread sampled at 5\,GS/s instead of 1\,GS/s in LST-1, no electronic noise, and a
background of protons only, simulated within $4.25^{\circ}$ of the pointing direction
and 250\,m of the telescope. Each of these idealisations favours the \svt\
(Sect.~\ref{sec:realism}). In one respect, the comparison disfavours the \svt: the cuts
of LST-1 are optimised for one hour (Appendix~\ref{app:ref}), whereas those of the
\svt\ are fixed (Sect.~\ref{sec:fom}). Section~\ref{sec:scan_pulse} shows how the performance
changes with the pulse width and the sampling rate.

\paragraph{Better photodetectors.} If the QE of the photodetectors improves, the
\svt\ comes closer to LST-1 and MAGIC. With a uniform QE of 50\,\%, its
energy threshold is 0.11\,TeV, instead of 0.18\,TeV with the LST-1 PMTs. It is then
2.1--2.5 times less sensitive than LST-1 between 0.1 and 0.63\,TeV, and 1.4--1.8
times between 0.63 and 6.3\,TeV. With photodetectors of 100\,\% QE, which no real
photodetector reaches, the threshold is 0.074\,TeV, and the \svt\ is 1.1--1.5 times
less sensitive than LST-1 between 0.1 and 6.3\,TeV. We adopt the LST-1 PMT response for the rest of this paper.

\subsection{Signal used by each analysis}
\label{sec:qe_pe}

\begin{figure}[t]
\centering
\includegraphics[width=\columnwidth]{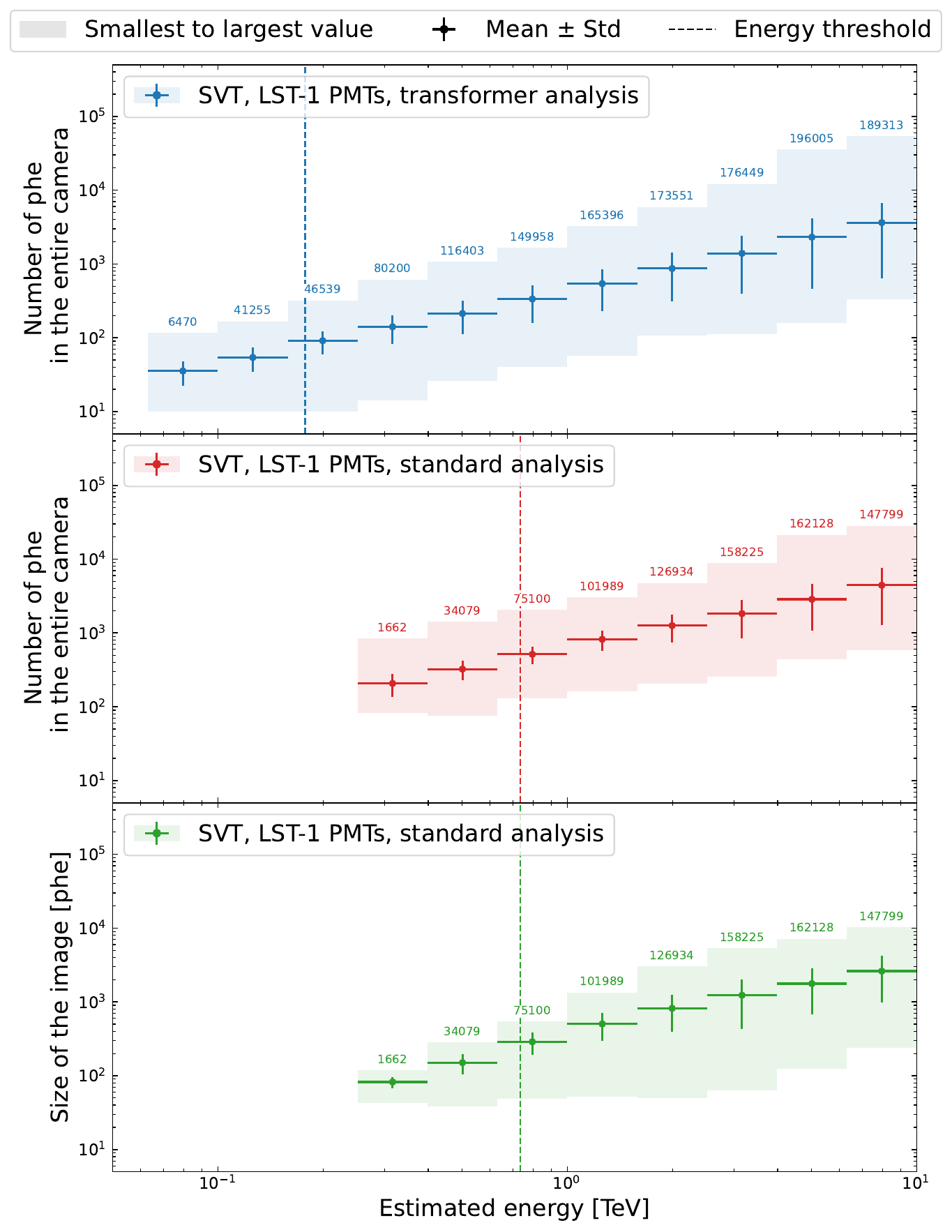}
\caption{Number of photoelectrons (phe) per simulated gamma-ray event versus
estimated energy, for the baseline telescope with the LST-1 PMTs. Top: phe
produced by the Cherenkov light in the whole camera, for the gamma rays that survive
the cuts of the transformer analysis. Middle: the same for the gamma rays that
survive the cuts of the standard analysis. Bottom: image size of the same events,
the sum of the charges of the pixels that survive the image cleaning, which is the
signal used by the standard analysis. Markers: mean; error bars: standard deviation;
shaded boxes: range from the smallest to the largest value. The number above each box
is the number of simulated events that survive the cuts in that bin. Dashed lines:
energy threshold of each analysis.}
\label{fig:pe}
\end{figure}

Figure~\ref{fig:pe} shows why the transformer reaches lower energies. It is produced for
the baseline telescope with the LST-1 PMTs, with the simulated gamma rays that
survive the cuts of each analysis. In each bin of estimated energy, it gives the
number of photoelectrons produced by the Cherenkov light in the whole camera, for the
events kept by the transformer analysis (top panel) and by the standard analysis
(middle panel). For the events of the standard analysis, the bottom panel shows the
image size, the sum of the charges of the pixels that survive the image cleaning.
The Hillas parameters, and hence the whole standard analysis, are computed from these
images.

The two analyses use the signal very differently. The calibration of the standard
analysis integrates the charge of each pixel in a short time window around the peak
of the pulse. The image cleaning then keeps only the pixels with enough charge, and
those of their neighbours whose charge exceeds a lower threshold. The image size is
therefore only 40--70\,\% of the number of photoelectrons in the camera (middle and
bottom panels; on their logarithmic scale the difference looks small). The transformer, instead,
reads the full waveform of every pixel. In its lowest energy bin, 0.063--0.1\,TeV,
the events have about 35 photoelectrons in the camera on average. The lowest bin of
the standard analysis, 0.25--0.4\,TeV, contains events with about 200 photoelectrons
in the camera, and their images have a size of about 80 photoelectrons. The standard
analysis thus needs events of about 200 photoelectrons in the camera, of which only
about 80 remain in the image after cleaning. The transformer can work with events of
only about 35 photoelectrons in the whole camera, about six times fewer. This is the
main reason why the energy threshold of the transformer analysis is four times lower
(Sect.~\ref{sec:qe_sens}). Stereoscopic systems can use fainter images,
because they combine several images of the same shower. MAGIC, for example,
typically requires an image size of at least 50 photoelectrons in each of its two
telescopes \citep{aleksic2016}.

\section{Dependence on the hardware parameters}
\label{sec:scan}

Section~\ref{sec:qe} gives the sensitivity of the baseline telescope with the
photomultipliers of LST-1. We now study how the performance of the transformer
analysis changes with the hardware. Starting from this baseline, we change one
parameter at a time: the dish diameter (Sect.~\ref{sec:scan_dish}) and the field of
view of the camera (Sect.~\ref{sec:scan_fov}). We then change the width of the photodetector pulse and the sampling rate of the
waveforms together (Sect.~\ref{sec:scan_pulse}). Table~\ref{tab:configs} lists the eight
configurations. The aim is to find which parameters matter for a transformer
analysis, not to find the best telescope configuration. Many more configurations
remain to be studied, and a dedicated optimisation will follow
(Sect.~\ref{sec:conclusions}).

Each configuration is simulated with \texttt{sim\_telarray} on the common set of
showers (Sect.~\ref{sec:mc}), and the transformer is trained and evaluated on it.
The hyperparameters are those of Paper~1 for every configuration
(Sect.~\ref{sec:vivit}). They were not optimised for the baseline, so the
comparison does not favour it. We do not interpret differences smaller than about 12\,\%, the largest difference
found between trainings with different random seeds (Sect.~\ref{sec:vivit}).
Figures~\ref{fig:dish}, \ref{fig:fov} and \ref{fig:pulse} show the same baseline
curve, which is also the transformer curve with the LST-1 PMTs of Fig.~\ref{fig:qe}.
Training the transformer for one configuration took about 7.5\,h of wall-clock time on
four NVIDIA A100 GPUs. This time varies by up to a factor of two between
configurations. With four GPUs, the training is faster than in Paper~1, which used a
single NVIDIA H200 GPU.

\begin{table}[t]
\caption{Configurations of Sect.~\ref{sec:scan}, all with the LST-1 PMTs, pixels of
$0.1^{\circ}$ and a focal ratio of 1.5, and their energy thresholds. The baseline is
in bold.}
\label{tab:configs}
\centering
\small\setlength{\tabcolsep}{3.5pt}
\begin{tabular}{l c c c c c}
\toprule
 & Dish & Field & Pulse & Sampling & $E_{\rm th}$ \\
 & diameter & of view & FWHM & rate & \\
 & [m] & [deg] & [ns] & [GS/s] & [TeV] \\
\midrule
\textbf{Baseline} & \textbf{5} & \textbf{6} & \textbf{1} & \textbf{5} & \textbf{0.18} \\
\midrule
Dish          & 4 & 6 & 1 & 5 & 0.38 \\
              & 6 & 6 & 1 & 5 & 0.13 \\
\midrule
Field of view & 5 & 4 & 1 & 5 & 0.26 \\
\midrule
Electronics\tablefootmark{a} & 5 & 6 & 2 & 5 & 0.18 \\
              & 5 & 6 & 3 & 5 & 0.19 \\
              & 5 & 6 & 2 & 2 & 0.26 \\
              & 5 & 6 & 3 & 2 & 0.27 \\
\bottomrule
\end{tabular}
\tablefoot{\tablefoottext{a}{Width of the single-photoelectron pulse, and sampling rate. A pulse of 1\,ns sampled
at 2\,GS/s is not studied, because it would have only two samples within its FWHM.}}
\end{table}

\subsection{Dish diameter}
\label{sec:scan_dish}

We use dish diameters of 4, 5 and 6\,m, with the focal ratio of 1.5 and
the camera of the baseline. The mirror area, and with it the number of Cherenkov
photons collected, changes by factors of 0.64 and 1.44 relative to the baseline. The
rate of night-sky photons in a pixel is
\begin{equation}
R \;=\; \frac{\pi}{4}\,B\,\varepsilon\,D^{2}\,\theta_{\rm pix}^{2},
\label{eq:nsb}
\end{equation}
where $\varepsilon$ is the photodetection efficiency and $B$ the sky brightness. Paper~1 used the
CTAO reference dark sky, $B=0.24$ photons\,sr$^{-1}$\,cm$^{-2}$\,ns$^{-1}$ between
300 and 650\,nm. Equation~(\ref{eq:nsb}) is Eq.~(1) of Paper~1, written with the
angular pixel size $\theta_{\rm pix}=d_{\rm pix}/f$. For the baseline, whose pixels
subtend $0.102^{\circ}$, it gives 0.15 photons per nanosecond for a QE of 100\,\%, the
value used in Paper~1. At a fixed angular pixel size, the night-sky rate in a pixel
changes by the same factors as the mirror area. Both the Cherenkov signal and the night-sky rate in a pixel thus grow as $D^{2}$, but
the fluctuations of the night-sky background grow only as $D$.

\begin{figure}[t]
\centering
\includegraphics[width=\hsize]{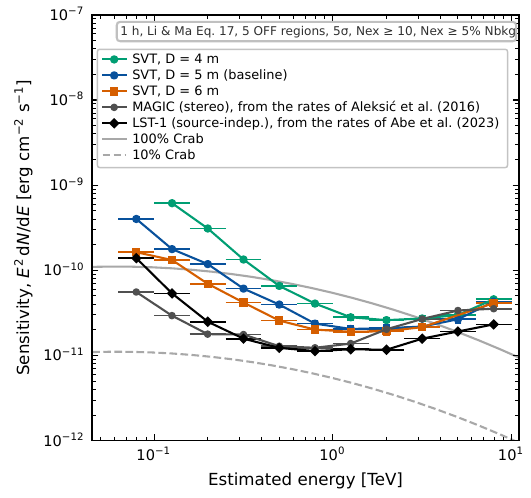}
\caption{Differential sensitivity in one hour of the telescope with dishes of 4\,m,
5\,m (baseline) and 6\,m, with the LST-1 PMTs and the
transformer analysis, compared with LST-1 and MAGIC as in Fig.~\ref{fig:qe}. A
larger dish lowers the energy threshold and improves the sensitivity at low
energies.}
\label{fig:dish}
\end{figure}

Figure~\ref{fig:dish} shows that a larger dish lowers the energy threshold and
improves the sensitivity at low energies. The threshold falls from 0.38\,TeV for a
4\,m dish to 0.18\,TeV for the baseline and 0.13\,TeV for a 6\,m dish. Between 5 and
6\,m, it falls by a factor of 1.40, close to the increase of the mirror area, 1.44.
Between 5 and 4\,m, it rises by a factor of 2.2, more than the mirror area
decreases, by a factor of 1.56. Compared with the baseline, the 6\,m dish is
1.3--1.7 times more sensitive between 0.1 and 0.63\,TeV, and the 4\,m dish 1.7--3.5
times less sensitive between 0.1 and 1\,TeV. The 6\,m dish also reaches lower
energies: between 0.063 and 0.1\,TeV, it is 2.5 times more sensitive than the
baseline, whereas the 4\,m dish does not reach this bin. Above 1\,TeV, where the
images are bright for all three dishes, the differences are smaller. The 6\,m dish
is then within 10\,\% of the baseline, and the 4\,m dish 1.1--1.4 times less
sensitive.

The analysis matters as much as the dish diameter. The 4\,m dish with the transformer
is more sensitive than the 5\,m dish with the standard analysis
(Sect.~\ref{sec:qe_sens}) in every energy bin in which both give a sensitivity: 10
times at 0.5\,TeV, and 1.2--1.8 times between 0.63 and 4\,TeV. Its energy threshold,
0.38\,TeV, is about half that of the standard analysis with the 5\,m dish,
0.73\,TeV.

\subsection{Field of view}
\label{sec:scan_fov}

The camera of the baseline has $60\times60$ pixels of $0.1^{\circ}$, a field of view
of $6^{\circ}$ on a side. We compare it with a camera of $40\times40$ such pixels,
with a field of view of $4^{\circ}$. The smaller camera has 1600 instead of 3600
channels.

\begin{figure}[t]
\centering
\includegraphics[width=\hsize]{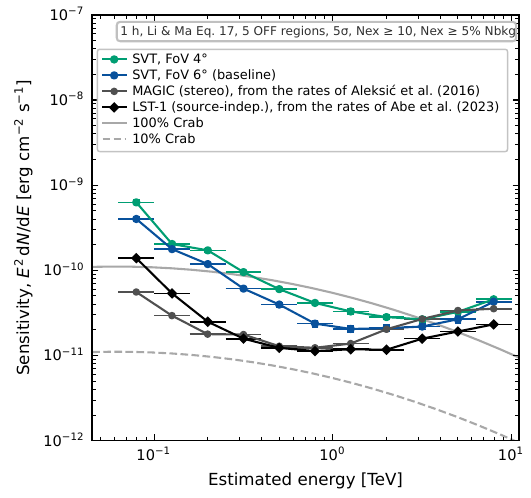}
\caption{As Fig.~\ref{fig:dish}, for cameras with fields of view of $4^{\circ}$ and
$6^{\circ}$ (baseline). The wider camera rejects the
background better and has a lower energy threshold.}
\label{fig:fov}
\end{figure}

Figure~\ref{fig:fov} shows that the $4^{\circ}$ camera is 1.2--1.8 times less
sensitive than the baseline between 0.1 and 1.6\,TeV, and 1.2--1.35 times less
sensitive between 1.6 and 6.3\,TeV. Its energy threshold is 0.26\,TeV, instead of
0.18\,TeV. With the smaller camera, more background events survive the cuts. Between 0.63
and 6.3\,TeV, the rate of protons after cuts is 2.5--5.4 times higher than with the
baseline, and the difference grows with energy. This has two causes. First, the
angular resolution is poorer. The radius that contains 68\,\% of the gamma rays,
which sets the cut on the angular distance to the source (Sect.~\ref{sec:fom}), is
1.2--1.5 times larger between 0.16 and 6.3\,TeV. The ON region is therefore 1.5--2.3 times larger, and so is the background in it. Second, the background rejection is
poorer. For the same cut on the gamma/hadron score, the smaller camera lets through
1.3--3.6 times more protons per unit solid angle between 0.63 and 6.3\,TeV, and this
factor also grows with energy. Between 0.1 and 0.63\,TeV, it lets through fewer
protons per unit solid angle, but it also keeps 25--44\,\% fewer gamma rays.
This is because the cut on the score is fixed (Sect.~\ref{sec:fom}), so that the two
cameras work at different gamma-ray efficiencies.

A wider camera rejects the background better because it records more of each proton
shower. When a cosmic-ray proton interacts in the atmosphere, it produces neutral
pions ($\pi^{0}$), which decay into pairs of photons, and each photon starts an
electromagnetic sub-shower. A proton shower looks like a gamma ray mainly when most
of its recorded light comes from a single sub-shower, or from the two sub-showers of
one $\pi^{0}$. In simulations of CTAO, such events make up at least half of the
background that survives the cuts below about 100\,GeV \citep{sitarek2018}. An event
is recognised as a hadronic shower when the light of further sub-showers, or of
muons, is also recorded. The secondary particles of the interactions receive a
transverse momentum, so that sub-showers can develop several degrees away from the
direction of the proton. Their Cherenkov light then reaches the camera a few degrees
away from the main image. A wider camera records more of this light, which helps the
network to reject the event and improves the sensitivity.

A wider field of view also helps to observe transient sources whose position is
poorly known. The $6^{\circ}$ camera covers 37\,deg$^{2}$ of the sky, 2.25 times more
than the $4^{\circ}$ camera. More of the error region of an alert can then be
observed in a single pointing. This is a real advantage, because many alerts are
poorly localised: the error regions of many gamma-ray bursts detected by the
Gamma-ray Burst Monitor on board \textit{Fermi} are several degrees wide
\citep{meegan2009,connaughton2015}, as Paper~3 discusses.

\subsection{Pulse width and sampling rate}
\label{sec:scan_pulse}

The width of the single-photoelectron pulse and the sampling rate determine how
finely the arrival times of the
photons are recorded, and how much the pulses of night-sky photons overlap with those
of the Cherenkov photons. We use pulse widths of 1\,ns, as in Paper~1, and of 2 and
3\,ns. We use sampling rates of 5\,GS/s, as in Paper~1, and of 2\,GS/s. The
multiplexed readout of MAGIC \citep{bartko2005}, installed in 2007, sampled at
2\,GS/s and recorded pulses of 2.3\,ns FWHM \citep{goebel2008}. We require at least
three samples within the FWHM of the single-photoelectron pulse. Of the six
combinations of three pulse widths and two sampling rates, this excludes a 1\,ns
pulse sampled at 2\,GS/s, which has only two samples within the FWHM, and leaves
five.
The recorded window is kept at about the 25.6\,ns of Paper~1: 128 samples at
5\,GS/s, and 52 samples (26\,ns) at 2\,GS/s.

\begin{figure}[t]
\centering
\includegraphics[width=\hsize]{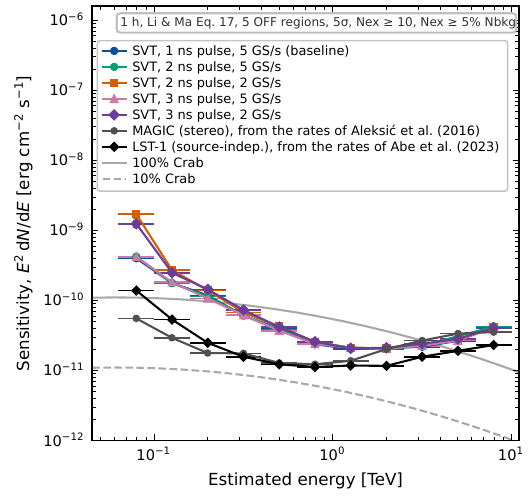}
\caption{As Fig.~\ref{fig:dish}, for single-photoelectron pulses of 1, 2 and 3\,ns FWHM,
sampled at 5 or 2\,GS/s (Table~\ref{tab:configs}). The pulse width matters little,
whereas sampling at 5\,GS/s lowers the energy threshold and improves the sensitivity
below 0.63\,TeV.}
\label{fig:pulse}
\end{figure}

Figure~\ref{fig:pulse} shows that the results depend little on the pulse width. At
5\,GS/s, pulses of 1, 2 and 3\,ns give energy thresholds of 0.18--0.19\,TeV. Their
sensitivities differ by at most 11\,\% at any energy, which is comparable with
the statistical uncertainties and with the variations between trainings
(Sect.~\ref{sec:vivit}). The sampling rate matters more. At 2\,GS/s, the
energy threshold rises to 0.26--0.27\,TeV, and the sensitivity degrades at low
energies: by a factor of 3--4 between 0.063 and 0.1\,TeV, 1.4--1.5 between 0.1 and
0.16\,TeV, and 1.1--1.35 between 0.16 and 0.63\,TeV. Above 0.63\,TeV, the sampling
rate changes the sensitivity by less than 10\,\%. For a transformer analysis, a fast
digitisation of the waveforms is therefore worth more than a narrow photodetector
pulse. This widens the choice of hardware: photodetectors and readout electronics that
give pulses up to 3\,ns wide can be used with almost no loss of performance.

\section{Discussion}
\label{sec:discussion}

The central result of Sect.~\ref{sec:qe} is that, for a given mirror area, the
performance of a compact telescope analysed with a transformer depends much less on
the photodetector efficiency than it does with the standard analysis. The main
reason is that the transformer can work with much fainter events
(Sect.~\ref{sec:qe_pe}).
Section~\ref{sec:scan} shows which hardware parameters matter for the transformer.
The dish diameter sets the energy threshold and the sensitivity below 1\,TeV. The
analysis matters as much: a 4\,m dish with the transformer has a lower energy threshold
than the 5\,m dish with the standard analysis, and it is more sensitive at all energies
(Sect.~\ref{sec:scan_dish}). The field of view affects the sensitivity over the whole energy range, through the
angular resolution and the rejection of the background. A wide field also helps to
observe poorly localised transients. The sampling rate matters below 0.63\,TeV, whereas the pulse
width, between 1 and 3\,ns, hardly matters.

Several limitations remain. The results come from simulations. A network trained
on simulations depends on the accuracy of the simulated instrument response more
than a parameter-based analysis does, precisely because it uses more of the information in the data
(Paper~1). A real telescope of this kind will therefore need a very accurate
simulation of its response (Paper~3). The simulated telescope also keeps the
idealisations of Sect.~\ref{sec:realism}, such as a perfectly reflecting mirror and
no electronic noise. We do not yet know whether they change the differences between
the configurations of Sect.~\ref{sec:scan}. Two effects act on the absolute
sensitivities in opposite directions. On the one hand, the idealisations of the
simulated telescope make the present values optimistic. On the other hand,
machine-learning methods and telescope hardware, such as photodetectors, progress
quickly. A telescope built a few years from now will therefore perform better.

The parameter study of Sect.~\ref{sec:scan} has limits too. It varies a few
parameters around the baseline. It uses the same network for every configuration,
with hyperparameters that were not optimised, and a fixed cut on the gamma/hadron score (Sects.~\ref{sec:vivit} and
\ref{sec:fom}). It shows which parameters
matter, but it does not identify the best telescope configuration, which is not a goal of this
study. The parameters may also
interact. For example, at a fixed angular pixel size, the night-sky background in a
pixel grows as the square of the dish diameter (Eq.~\ref{eq:nsb}), and its effect
depends on the sampling rate. A joint optimisation of telescope and
network is beyond the scope of this paper.

\section{Conclusions and outlook}
\label{sec:conclusions}

This paper builds on Paper~1, which showed in simulation that a video vision
transformer lowers the energy threshold of an idealised 5\,m telescope by a factor
of three. We replaced the photodetectors of Paper~1, with a QE of 100\,\%, by
realistic ones, and analysed the same simulated events with the transformer and with
the standard analysis. We computed the differential flux sensitivity in one hour,
which Paper~1 did not report, and compared it with that of LST-1 and MAGIC. We then
studied how the dish diameter, the field of view, the pulse width and the sampling
rate affect the transformer analysis.

The performance of a stand-alone telescope with a 5\,m diameter mirror depends far less on the
photodetector efficiency when its data are analysed with a video vision transformer
than with the standard analysis. With the same photodetectors, the transformer
analysis is more sensitive at all energies, and most of all at low energies. With
the photomultipliers of LST-1, it is 17 times more sensitive than the standard
analysis at 0.5\,TeV and 1.5 times above 2.5\,TeV, and its energy threshold is four times lower. The main
reason is that the transformer can work with events of about 35 photoelectrons in the
camera, whereas the standard analysis needs events of about 200 photoelectrons, of
which only about 80 remain in the image after cleaning. The change of analysis is worth more than a 100\,\% QE photodetector at low
energies. With the photomultipliers of LST-1, the transformer analysis is more
sensitive below 0.4\,TeV than the standard analysis with photodetectors of 100\,\%
QE, and it has about the same energy threshold. The change of analysis is also worth
more than an increase of the dish diameter from 4 to 5\,m. A 4\,m dish with the
transformer is more sensitive at all energies than the 5\,m dish with the standard
analysis, and its energy threshold is about half as high, 0.38 instead of 0.73\,TeV.

In a simulation that is still
idealised (Sect.~\ref{sec:realism}), the 5\,m telescope with these photomultipliers is
3.2--4.8 times less sensitive in one hour than the 23\,m LST-1 between 0.1
and 0.63\,TeV, and 1.4--2.1 times less sensitive between 0.63 and 6.3\,TeV. Better
photodetectors would bring it closer to LST-1 and MAGIC: with a QE of 50\,\%, it
would be 2.1--2.5 and 1.4--1.8 times less sensitive than LST-1 in the same energy
ranges. With the transformer
analysis, a larger dish lowers the energy threshold and improves the sensitivity
below 1\,TeV. A $6^{\circ}$ camera rejects the background better than a $4^{\circ}$
camera. Sampling at 5\,GS/s improves the performance below 0.63\,TeV, whereas the
pulse width matters little.

Together with Paper~1, this is the first application of a transformer analysis to a
compact IACT, and it is only a beginning. Already at this early stage, the transformer lowers the energy
threshold by a factor of four compared with the standard analysis, and it is
3--17 times more sensitive between 0.4 and 1\,TeV (Sect.~\ref{sec:qe_sens}). We are confident
that it will improve further, as we learn to optimise such analyses and as more
powerful methods become available (see below). If a compact telescope with a
transformer analysis is built in the future, its hardware will also be better than
that available today. For instance, the PMTs of LST-1 were produced in 2015, and
photodetectors keep improving. With a QE of 50\,\%, the energy threshold of the
baseline would fall from 0.18 to 0.11\,TeV
(Sect.~\ref{sec:qe_sens}). The study can be extended in three directions.

\paragraph{Improving the analysis.} The analysis of this paper is deliberately
simple: the cut on the gamma/hadron score is fixed, and the network is the same
for every telescope configuration. Optimising the cuts in each energy bin should improve the
sensitivity by up to a few tens of per cent (Sect.~\ref{sec:fom}), in particular at the
lowest energies (Sect.~\ref{sec:mc}). An initial scan in Paper~1 showed that the maximum
learning rate, the embedding dimension and the relative weight of the regression term
in the composite loss are the hyperparameters that matter most. Optimising them for
each telescope configuration should improve the performance. At the lowest energies, the night-sky background
limits the performance. A network trained to remove it, applied before the
reconstruction, could improve the performance there (Paper~1). At the highest
energies, a larger sample of simulated protons would give the sensitivity in every
bin: above a few TeV, fewer than four simulated protons survive the cuts of some
configurations (Sect.~\ref{sec:fom}). Two developments of machine learning are also
promising. Networks first trained on large and varied datasets, and then adapted to a
specific task, can reach a given accuracy with far fewer examples of that task
\citep{subramanian2023}. Such a network could be pre-trained on the large archives of
real data of current IACTs. The performance of networks also improves predictably as
the network and its training dataset grow, as found for the classification of jets in
particle physics \citep{vigl2026}. The network itself can be improved in several ways. Its
input could be cut into tokens in a way designed for Cherenkov data, instead of the
regular tubelets used here (Sect.~\ref{sec:vivit}). It could start from a pre-trained
vision backbone, such as DINOv3 \citep{simeoni2025}. Such a backbone is the part of a
network that extracts the features of an image, and it has been trained on very large
sets of images. The network could exploit the sparsity of the raw data, in which most
pixels contain only night-sky light, to reduce the computing cost. It could also make
use of the symmetries of the Cherenkov signals. Finally, transformers, and the methods
to train them, develop quickly, and more powerful networks will become available.

\paragraph{Other telescope configurations.} This paper varies four parameters around
the baseline. Many more configurations will be studied: other pixel sizes, focal
ratios, mirror designs and fields of view, and simultaneous changes of several parameters, with the
more realistic telescope described below. Such a study is not a simple repetition of
the present one, because the transformer must be adapted to each configuration. A
camera with the same field of view but smaller pixels, for example, has more channels, and the network then
needs more trainable parameters. Its architecture, its size and its hyperparameters must therefore be adjusted
for each configuration, and the training time grows with the size of the input.
In this paper, the same network is used for every configuration
(Sect.~\ref{sec:vivit}), which is why the pixel size is not varied and the field
of view is not enlarged. Instead of a predefined
set of configurations, a future study could use an optimisation algorithm based on
artificial intelligence. Such algorithms have already been used to design
instruments. Bayesian optimisation improved the design of a ring-imaging Cherenkov
detector for the future Electron-Ion Collider \citep{cisbani2020}. A search driven
by artificial intelligence found interferometer layouts for gravitational-wave
detectors that, under realistic constraints, outperform the designs planned for the
next generation \citep{krenn2025}. Differentiable programming enables the
end-to-end optimisation of particle-physics instruments \citep{dorigo2023}, and it
has been applied to the layout of the Southern Wide-field Gamma-ray Observatory
\citep{dorigo2025}. We intend to apply such methods to the telescope, jointly with
the hyperparameters of the network. Each step of such an optimisation needs
simulated events for a new configuration. Generative networks trained on the full
simulation can produce them much faster than the simulation itself. Such networks
have been developed for particle showers in calorimeters \citep{paganini2018} and
for the images of imaging atmospheric Cherenkov telescopes, which they produce
about $10^{5}$ times faster than the full simulation \citep{elflein2024}.

\paragraph{A more realistic telescope.} This is a first study of a transformer
analysis for a compact IACT, and we kept it simple on purpose. The simulated
telescope therefore keeps several idealisations of Paper~1 (Sect.~\ref{sec:realism}).
Now that the results show that a large improvement is possible, the next step is to
make the simulations, and the whole study, more realistic. A future paper will study
these idealisations on the baseline: several levels of night-sky background, up to
those of moderate moonlight; a realistic mirror reflectivity, with facet errors and
misalignment added to the ray tracing; the transit-time spread of the
photodetectors; and the electronic noise. It will also simulate protons at larger
angles from the pointing direction and at larger distances from the telescope. Such
simulations are less efficient, because many more of the simulated protons do not
trigger the telescope, but they are closer to real operation. Future studies should also report the performance of transformer
analyses at zenith angles up to $70^{\circ}$. Such zenith angles matter if one wants to
increase the temporal coverage of a source (Paper~3).

\begin{acknowledgements}
\textit{Author contributions.} E.~Jobst: main developer of the transformer analysis;
Monte Carlo simulations of the telescope configurations; training and validation of
the transformer analysis and of the standard analysis for these configurations;
writing -- review and editing. D.~Paneque: conceptualisation of the contents of the
paper; supervision; computation and visualisation of the flux sensitivities; project
administration; funding acquisition; writing -- original draft, review and editing.
L.~Heinrich: methodology and supervision of the transformer analysis; funding
acquisition; writing -- review and editing.

We are very grateful to R.~Mirzoyan, who provided a detailed and constructive
list of remarks at short notice, on the last day before submission. We also
thank A.~Berti, A.~Moralejo and R.~L\'opez-Coto for discussions and for
information on the sensitivity of LST-1. We are grateful to the developers of
the CTAO \texttt{simtools} package, a reliable and easy-to-use code that we used
extensively for our simulations. The computations were performed on the
high-performance computing clusters of the Max Planck Computing and Data Facility
(MPCDF) and of the Origins Data Science Lab (ODSL). D.P. acknowledges support from
the Deutsche Forschungsgemeinschaft (DFG, German Research Foundation) under Germany's
Excellence Strategy -- EXC-2094 -- 390783311. E.J. and L.H. are supported by the BMFTR project
SciFM (05D25WO2).
\end{acknowledgements}

\bibliographystyle{aa}
\bibliography{references}

\begin{appendix}

\section{Sensitivities of MAGIC and LST-1 in one hour}
\label{app:ref}

The MAGIC and LST-1 collaborations report the differential sensitivity for 50\,h of
observation \citep{aleksic2016,abe2023}. This appendix derives the sensitivities in
1\,h used in this paper (Sect.~\ref{sec:fom}). We compute them from the rates of
gamma rays and background after cuts, in five bins per decade of estimated energy,
for a source with the spectrum of the Crab Nebula. The number of events expected in
an observation is the rate multiplied by its duration. The sensitivity is the flux
that gives a $5\sigma$ excess according to Eq.~(17) of \citet{lima1983}, with five
OFF regions, at least ten excess events, and an excess of at least 5\,\% of the
background. \citet{abe2023} used this definition for LST-1, and we use it for the
\svt. All the curves start in the bin of 0.063--0.1\,TeV.

\begin{figure*}[t]
\centering
\includegraphics[width=\textwidth]{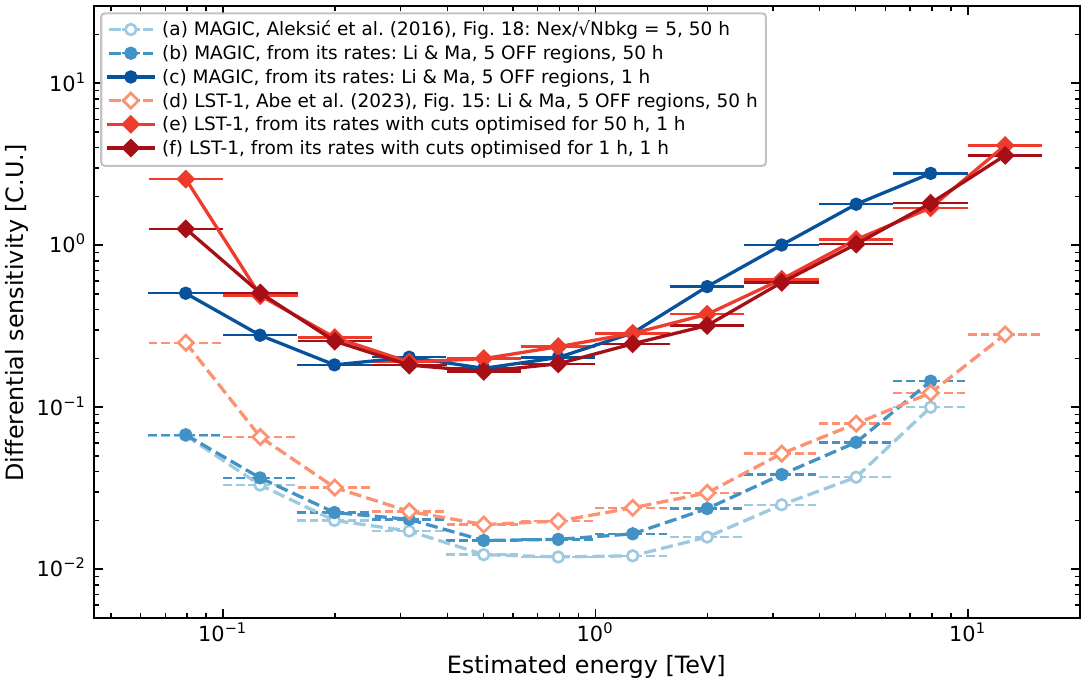}
\caption{Differential sensitivity of MAGIC and LST-1 in Crab units (C.U.), in five
bins per decade of estimated energy. Dashed lines: 50\,h; solid lines: 1\,h. Open
markers: published curves; filled markers: curves computed from the rates of gamma
rays and background (Appendix~\ref{app:ref}). Curves (c) and (f) are the references
used for the \svt\ in this paper.}
\label{fig:ref}
\end{figure*}

Figure~\ref{fig:ref} shows six curves. Curve (a) is the published sensitivity of
MAGIC for 50\,h, the red points of Fig.~18 of \citet{aleksic2016}. It uses a simpler
definition of the significance. The sensitivity is the flux for which the number of
excess events divided by the square root of the number of background events,
$N_{\rm ex}/\sqrt{N_{\rm bkg}}$, equals 5. The two other conditions are the same
as in our definition: at least ten excess events, and an excess of at least 5\,\% of
the background. Curve (b) is MAGIC for 50\,h, computed from the rates of Table~A.7 of
\citet{aleksic2016}. In every bin, it reproduces the values given in that table for
Eq.~(17) with five OFF regions within the rounding of the published numbers
(differences of at most 3.4\,\%). Curve (c) is the same for 1\,h.
Curve (d) is the published sensitivity of LST-1 for 50\,h, from the
source-independent analysis of \citet[Fig.~15]{abe2023}. Curves (e) and (f) are
LST-1 for 1\,h, computed from the rates of the same analysis. For curve (e), the cuts
were optimised for 50\,h. For curve (f), the cuts were optimised for 1\,h. Both
sets of rates were provided by the LST collaboration.

We compare the curves in the bins of 0.16--0.25\,TeV and 1.0--1.6\,TeV. For MAGIC in
50\,h, the sensitivity with Eq.~(17) and five OFF regions is worse by factors of
1.12 and 1.37 than with $N_{\rm ex}/\sqrt{N_{\rm bkg}}=5$ (curves b and a). The simpler
definition neglects the statistical fluctuations of the background estimate and of
the signal, which matter most when the background is low. From 50\,h to 1\,h, the
sensitivity of MAGIC degrades by factors of 8.2 and 17 (curves c and b). That of
LST-1, computed from the same rates with the cuts optimised for 50\,h, degrades by
factors of 8.0 and 13. These factors are larger than $\sqrt{50}\approx7.1$, the
factor expected when the background dominates. In 1\,h, few background events remain
above a few hundred GeV, and the sensitivity is then limited by the small number of
excess events. For LST-1 in 1\,h, the sensitivity with the cuts optimised for 50\,h
is worse by factors of 1.05 and 1.16 than with the cuts optimised for 1\,h (curves e
and f). Above
0.1\,TeV, the two differ by less than 30\,\%, and in the bin of 0.063--0.1\,TeV by a
factor of two.

The references used for the \svt\ in this paper are curve (c) for MAGIC, with the
cuts optimised for 50\,h, and curve (f) for LST-1, with the cuts optimised for 1\,h.
No analysis of the MAGIC data with cuts optimised for 1\,h could be made in the time
available. We expect it to differ from curve (c) by amounts similar to those found
for LST-1, less than about 30\,\% above 0.1\,TeV.

\end{appendix}

\end{document}